\documentclass[twocolumn,showpacs,preprintnumbers]{revtex4}%
\usepackage{amssymb}
\usepackage{amsmath}
\usepackage{graphicx}
\usepackage{dcolumn}
\usepackage{bm}
\usepackage{amsfonts}%
\begin{document}
\title{Optomechanical Cavity with a Buckled Mirror}
\author{D. Yuvaraj}
\affiliation{Department of Electrical Engineering, Technion, Haifa 32000 Israel}
\author{M. B. Kadam }
\affiliation{Department of Electrical Engineering, Technion, Haifa 32000 Israel}
\author{Oleg Shtempluck }
\affiliation{Department of Electrical Engineering, Technion, Haifa 32000 Israel}
\author{Eyal Buks }
\affiliation{Department of Electrical Engineering, Technion, Haifa 32000 Israel}
\date{\today }

\begin{abstract}
We study an optomechanical cavity, in which a buckled suspended beam serves as
a mirror. The mechanical resonance frequency of the beam obtains a minimum
value near the buckling temperature. Contrary to the common case, in which
self-excited oscillations of the suspended mirror are optically induced by
injecting blue detuned laser light, in our case self-excited oscillations are
observed with red detuned light. These observations are attributed to a
retarded thermal (i.e. bolometric) force acting on the buckled mirror in the
inwards direction (i.e. towards to other mirror). With relatively high laser
power other interesting effects are observed including period doubling of
self-excited oscillations and intermode coupling.

\end{abstract}
\maketitle





\section{Introduction}

Systems combining mechanical elements in optical resonance cavities are
currently a subject of intense interest \cite{Braginsky&Manukin_67,
Hane_179,Gigan_67,Metzger_1002,Kippenberg_1172,Favero_104101}. Coupling of
nanomechanical mirror resonators to optical modes of high-finesse cavities
mediated by the radiation pressure has a promise of bringing the mechanical
resonators into the quantum realm \cite{Kimble_et_al_01, Carmon_et_al_05,
Arcizet_et_al_06, Gigan_67, Jayich_et_al_08, Schliesser_et_al_08,
Genes_et_al_08, Kippenberg_1172, Teufel_et_al_10} (see Ref. \cite{Poot_273}
for a recent review). In addition, the micro-optoelectromechanical systems
(MOEMS) are expected to play an increasing role in optical communications
\cite{Wu_et_al_06} and other photonics applications \cite{Stokes_et_al_90,
Lyshevski&Lyshevski_03, Hossein-Zadeh&Vahala_10}.

Besides the radiation pressure, another important force that contributes to
the optomechanical coupling in MOEMS is the bolometric force
\cite{Metzger_1002, Jourdan_et_al_08, Metzger_133903, Marino&Marin_10,
Metzger_133903, Restrepo_860, Liberato_et_al_10,Marquardt_103901,
Paternostro_et_al_06}, also known as the thermal force. This force can be
attributed to the heat induced deformations of the micromechanical mirrors
\cite{Aubin_et_al_04, Marquardt_103901, Paternostro_et_al_06,
Liberato_et_al_10}. In general, the thermal force plays an important role in
relatively large mirrors, in which the thermal relaxation rate is comparable
to the mechanical resonance frequency. Phenomena such as mode cooling and
self-excited oscillations
\cite{Hane_179,Kim_1454225,Aubin_1018,Carmon_223902,Marquardt_103901,Corbitt_021802,Carmon_123901,Metzger_133903}
have been shown in systems in which this force is
dominant~\cite{Metzger_133903, Metzger_1002, Aubin_et_al_04,
Jourdan_et_al_08,Zaitsev_046605,Zaitsev_1104_2235}.

In this paper we investigate an optomechanical cavity having a suspended
mirror in the shape of a trilayer metallic doubly clamped beam
\cite{Michael_870,Ross_537,Riethmuller_758}. The system experimentally
exhibits some unusual behaviors. For example, contrary to other experiments,
in which self-excited oscillations of the mechanical resonator are optically
induced when the cavity is blue detuned \cite{Rodrigues_053601}, here the same
effect occurs for the case of red-detuned cavity. Moreover, the dependence of
the mechanical resonance frequency on laser power is found to be non-monotonic
\cite{Okamoto_062202}. These findings are attributed to optically induced
thermal strain in the beam, which gives rise to compressive stress that may
result in buckling
\cite{Fang_116,Coffin_935,Gere_Mechanics_of_Mat,Talghader_R109,Ettouhami_167,McCarthy_37}%
. Bagheri \textit{et al}.. \cite{Bagheri_726} have recently reported the
utilization of the buckling phenomenon to develop a non-volatile mechanical
memory element in a similar optomechanical system (see also Ref.
\cite{Roodenburg_183501}).

We generalized the theoretical model that has been developed in Ref.
\cite{Zaitsev_1104_2235} to account for the effect of buckling
\cite{Nayfeh_1121,Lawrence_223}. We find that close to buckling the effective
thermal force acting on the beam can become very large, and consequently
optomechanical coupling is greatly enhanced. Consequently, the threshold laser
power needed for optically driving self-excited oscillations in the mirror can
be significantly reduced. We show that the proposed theoretical model can
account for some of the experimental results. On the other hand, some other
experimental observations remain elusive. For example, self-modulation of
self-excited oscillations is observed in some ranges of operation. Further
theoretical study, which will reveal the complex structure of stability zones
and bifurcations of the system \cite{Blocher_52}, is needed in order to
account for such findings.

\section{Experimental setup}

\begin{figure}
[ptb]
\begin{center}
\includegraphics[
height=3.7256in,
width=3.2396in
]%
{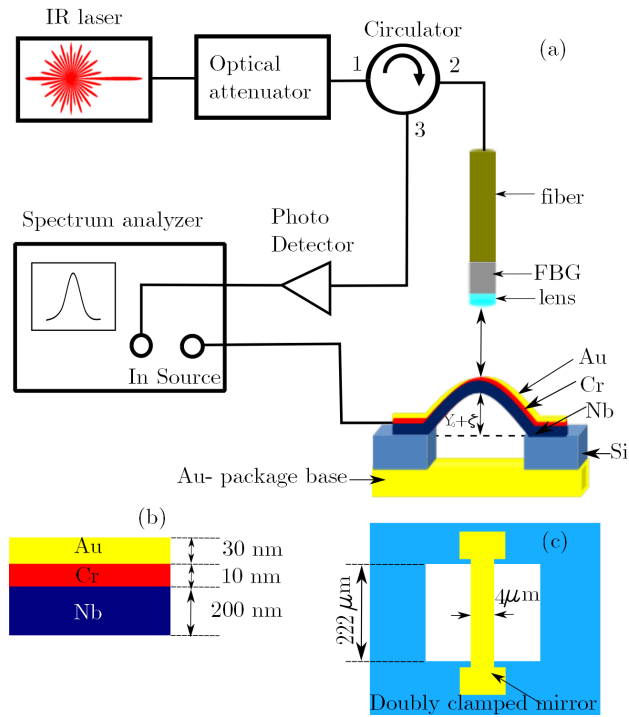}%
\caption{The experimental setup. (a) Sample and optical displacement sensor.
(b) Side view sketch of the trilayer beam mirror. (c) Top view of the
suspended trilayer beam A.}%
\label{Fig:setup}%
\end{center}
\end{figure}

The experimental setup (see Fig. \ref{Fig:setup}) is similar to the one
employed in Ref. \cite{Zaitsev_046605}. A fiber Bragg grating (FBG)
\cite{Snyder&Love_book_83,Snyder_Optical_Waveguide_Th,Poladian_2963} and a
microlens \cite{Mao_5887} are attached to the end of a single mode fiber. The
fiber can be accurately positioned using piezomotors. The system, which
consists of doubly clamped suspended multilayer micromechanical beam and the
optical fiber monitoring assembly, is located inside a cryostat maintained
under typical pressure of $10^{-3}%
\operatorname{mbar}%
$ and temperature of $77%
\operatorname{K}%
$. Optical cavity is formed between the freely suspended beam oscillating
parallel to the optical axis of the cavity and between either the FBG or the
glass-vacuum interface at the tip of the fiber, as shown in Fig.
\ref{Fig:setup}(a).

The multilayer beam is fabricated by using bulk micromachining process. A $200%
\operatorname{nm}%
$ thick Nb layer is coated on prefabricated silicon nitride membranes over
silicon substrates using DC magnetron sputtering at working pressure of
$5.2\times10^{-3}%
\operatorname{torr}%
$ and argon atmosphere. Patterning on the coated substrates is done using
photolithography followed by a liftoff process, in which a $10%
\operatorname{nm}%
$ thick layer of Cr and a $30%
\operatorname{nm}%
$ thick layer of gold-palladium (Au$_{0.85}$Pd$_{0.15}$) is patterned on top
of the Nb layer. Front electron cyclotron resonance (ECR) plasma etching of
the unmasked Nb and silicon nitride is done to suspend the AuPd/Cr/Nb
multilayer beam. Process is followed by a backside ECR\ plasma etching of the
suspended beam to remove silicon nitride form back side. Figure
\ref{Fig:setup}(b) depicts the sketch of the multilayer AuPd/Cr/Nb suspended
micromechanical beam having suspended length $l=222%
\operatorname{\mu m}%
$ , width $b=4%
\operatorname{\mu m}%
$ and total thickness $h=240%
\operatorname{nm}%
$ with\ effective Young's modulus $E=109.68\times10^{9}%
\operatorname{Pa}%
$, density $\rho=9715.22%
\operatorname{kg}%
/%
\operatorname{m}%
^{3}$, thermal expansion coefficient $\alpha=7.76\times10^{-6}%
\operatorname{K}%
^{-1}$ and Poisson's ratio $\sigma=0.3961$. These parameters are computed by
averaging over the layers using expressions taken from Ref. \cite{Ross_537}.
The top view of the suspended trilayer beam mirror that is described in the
following section is shown in Fig. \ref{Fig:setup}(c). We will refer to this
beam hereafter as beam A.%

The suspended multilayer beam can be capacitively actuated by applying voltage
to it with respect to the ground plate on which the sample is mounted. The
ground plate is located at $500%
\operatorname{\mu m}%
$ below the suspended beam. Incident optical power is controlled by
attenuating the injected laser power of wavelength $\lambda\simeq1550%
\operatorname{nm}%
$. The optical power reflected off the cavity is fed to a photodetector, which
is monitored using a spectrum analyzer, as shown in Fig. \ref{Fig:setup}(a).

\section{Self-Excited Oscillations Induced by Red-Detuned Cavity}

\begin{figure}
[ptb]
\begin{center}
\includegraphics[
height=2.5797in,
width=3.2396in
]%
{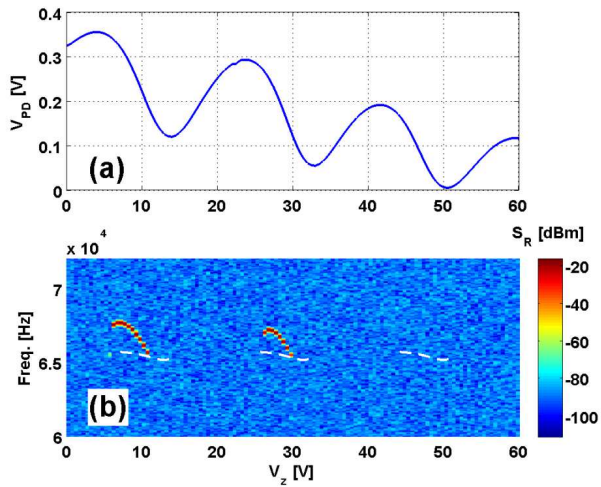}%
\caption{(Color Online) Dependence on cavity length (beam A). (a) The
photodetector DC voltage as a function of the voltage $V_{\mathrm{z}}$ applied
to the piezomotor holding the fiber. A change of $18.5\operatorname{V}$ in
$V_{\mathrm{z}}$ corresponds to a displacement of the fiber by $\lambda/2$.
The laser power in this measurement is set to the value of
$0.08\operatorname{mW}$, which is lower than the threshold of self-excited
oscillations. (b) Spectrum analyzer measurement of the photodetector voltage
measured with a laser power of $0.16\operatorname{mW}$. Self-excited
oscillations are observed when the cavity is red-detuned. The white dotted
line shows the calculated value of $\Omega_{\mathrm{eff}}/2\pi$ for the
regions in $V_{z}$ for which $\gamma_{\mathrm{eff}}<0$. The following
parameters have been used for the theoretical calculation $\theta_{\mathrm{C}%
}=0.06$, $\theta_{\mathrm{f}}=0.1$, $\Omega_{0}=1.2\times10^{-4}$,
$\xi_{\mathrm{C}}=2.8\times10^{-3}$, $\gamma=0.5\times10^{-9}$, $T_{\mathrm{B}%
}=0.96$, $T_{\mathrm{A}}=0.01$, $T_{\mathrm{R}}=0.6$, $\beta_{\mathrm{Y}%
}=-1600$ and $\beta_{\mathrm{TR}}=2.6\times10^{-7}$.}%
\label{Fig:S_R}%
\end{center}
\end{figure}

As was pointed out above, the optomechanical cavity under study exhibits some
unusual behaviors. As is demonstrated in Fig. \ref{Fig:S_R}, self-excited
oscillations can be induced by red-detuned cavity. Subplot (a) shows the
reflected optical power vs. the voltage $V_{\mathrm{z}}$, which is applied to
the piezomotor that is used to position the optical fiber along the optical
axis direction. The period of oscillation in the reflected power is
$\lambda/2$. In this experiment the wavelength $\lambda$ is not tuned to the
reflective band formed by the FBG, and thus the optical cavity is formed
between the freely suspended beam mirror and the glass-vacuum interface at the
tip of the fiber. For this case the finesse of the cavity is much lower
compared with values that can be achieved when the FBG is employed. Note that
in addition to the oscillatory part, also the averaged (over one period of
oscillation) measured reflected power exhibits dependence on cavity length
[see subplot \ref{Fig:S_R}(a)]. The average value obtains a maximum when the
cavity length coincides with the focal distance of the microlens. This occurs
close to the value $V_{\mathrm{z}}=0$.

While the data seen in subplot \ref{Fig:S_R}(a) is taken when the input
optical power is attenuated well below the threshold of self-excited
oscillations, in subplot \ref{Fig:S_R}(b), which depicts the measured spectrum
of reflected optical signal, the input power is set to $0.16%
\operatorname{mW}%
$. Self excited oscillations of frequency of about $67%
\operatorname{kHz}%
$ are found near $V_{\mathrm{z}}=8%
\operatorname{V}%
$ and near $V_{\mathrm{z}}=28%
\operatorname{V}%
$. In both these regions where self excited oscillations occur the
reflectivity decreases with $V_{\mathrm{z}}$, i.e. the cavity is effectively
red-detuned. As was discussed above, the deviation from periodic dependence of
the measured spectrum on cavity length with $\lambda/2$ period is attributed
to the effect of the microlens.%

\begin{figure}
[ptb]
\begin{center}
\includegraphics[
height=2.4396in,
width=3.2396in
]%
{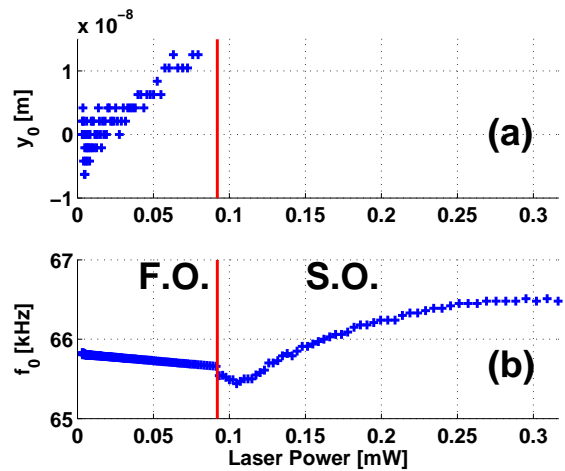}%
\caption{Dependence of laser power (beam A). (a) The static deflection $y_{0}%
$. (b) The mechanical frequency $f_{0}$. The vertical red line represents the
threshold of self-excited oscillations. Below the threshold $f_{0}$ is found
from the peak amplitude of forced oscillations, whereas above the threshold
$f_{0}$ is the frequency of self-excited oscillations.}%
\label{Fig:y_0_f_0}%
\end{center}
\end{figure}

The dependence on laser power and $P_{\mathrm{L}}$ is seen in Fig.
\ref{Fig:y_0_f_0}. The threshold power of self-excited oscillations occurs at
$0.092%
\operatorname{mW}%
$. In subplot \ref{Fig:y_0_f_0}(a) the static position $y_{0}$ of the center
of beam A is measured vs. laser power below the threshold power. This is done
by measuring the the reflected optical power vs. the voltage $V_{\mathrm{z}}$.
For each value of the laser power this yield a plot similar to the one see in
Fig. \ref{Fig:S_R}(a). However, the phase of oscillations is found to depend
on laser power. The value of the static deflection $y_{0}$, which is extracted
from that shift in the phase, is seen in subplot \ref{Fig:y_0_f_0}(a). The
observed behavior indicates that the beam is deflected towards the optical
fiber with increasing laser power. As will be seen below, this behavior
indicates that the thermal force in the present case acts in the inwards
direction, contrary to the more common situation (e.g. for the case of
radiation pressure), where cavity optomechanical forces act in the outwards
direction, and lead to elongation of the cavity (rather than shortening as in
the current case).

In subplot \ref{Fig:y_0_f_0}(b) the mechanical resonance frequency $f_{0}$ is
measured both below and above the threshold power of self-excited
oscillations. While below the threshold of $0.092%
\operatorname{mW}%
$ the frequency $f_{0}$ is determined from forced oscillations (FO) (i.e. from
the peak in the measured frequency response), above threshold $f_{0}$ is
determined from the peak in the spectrum of self-excited oscillations (SO).
Note the non-monotonic dependence of $f_{0}$ on laser power. As will be argued
below, this dependence suggests that the beam undergoes thermal buckling.%

\section{Mechanical Equations of Motion}

Most of the experimental findings that were presented in the previous section
are not accountable by the theoretical model that has been developed in Ref.
\cite{Zaitsev_1104_2235} (which, on the other hand, was very successful in
accounting for previous measurements of Ref. \cite{Zaitsev_046605}). We
therefore generalize the model to account for the effect of buckling in the
mirror. While the derivation of the system's equations of motion is described
in the appendices, the final equations are presented below in this section.

The height function $y\left(  x,t\right)  $ is assumed to have the shape of
the first buckling configuration of a doubly clamped beam
\cite{Nayfeh_1121,Lawrence_223}. Thus, for the case where the clamping points
are located at $\left(  x,y\right)  =\left(  \pm l/2,0\right)  $ the height
function $y\left(  x,t\right)  $ can be written as%
\begin{equation}
y=\xi l\left(  1+\cos\frac{2\pi x}{l}\right)  \;, \label{y 1st}%
\end{equation}
where $\xi$ depends on time. The equation of motion for $\xi$, which is
derived in appendix C, is given by%
\begin{equation}
\ddot{\xi}+2\gamma\dot{\xi}+\Omega^{2}\xi=F_{\mathrm{th}}+F_{\mathrm{p}%
}e^{-i\Omega_{\mathrm{p}}\tau}\;, \label{EOM xi}%
\end{equation}
where overdot denotes a derivative with respect to the dimensionless time
$\tau$, which is related to the time $t$ by the relation $\tau=t/\sqrt{\rho
A_{\mathrm{cs}}/E}$, where $\rho\ $is the mass density, $A_{\mathrm{cs}}$ is
the cross section area of the beam, $E$ is the Young's modulus, $\gamma$ is
the dimensionless damping constant, and $F_{\mathrm{p}}$ and $\Omega
_{\mathrm{p}}$ are the dimensionless amplitude and frequency respectively of
an externally applied force. The equation of motion (\ref{EOM xi}) contains
two thermo optomechanical terms that depend on the temperature of the beam
$T$. The first is the temperature dependent angular resonance frequency
$\Omega$ and the second is the thermal force $F_{\mathrm{th}}$. In terms of
the dimensionless temperature $\theta=\left(  T-T_{0}\right)  /T_{0}$, where
$T_{0}$ is the temperature of the supporting substrate (i.e. the base
temperature) the frequency $\Omega$ is given by $\Omega=\Omega_{0}\nu\left(
\theta\right)  $, where $\Omega_{0}=\sqrt{\rho A_{\mathrm{cs}}/E}\omega_{0}$,
$\omega_{0}$ is the mode's angular resonance frequency and the temperature
dependence is expressed in terms of the dimensionless function $\nu\left(
\theta\right)  $. The dimensionless thermal force is given by $F_{\mathrm{th}%
}=\Omega^{2}\xi_{\mathrm{C}}f_{\mathrm{Y}}\left(  \theta\right)  $, where
$\xi_{\mathrm{C}}=\Omega_{0}\sqrt{3T_{0}\left(  \alpha-\alpha_{\mathrm{s}%
}\right)  l^{4}/16\pi^{6}I}$, where $\alpha$ and $\alpha_{\mathrm{s}}$ are the
thermal expansion coefficients of the metallic beam and of the substrate
respectively, $I$ is the moment of inertia corresponding to bending in the
$xy$ plane, and the dimensionless function $f_{\mathrm{Y}}\left(
\theta\right)  $ represents the beam's temperature dependent deflection. For
the case of a rectangular cross section having width $l_{y}$ in the $y$
direction and width $l_{z}$ in the $z$ direction, and for the case of bending
in the $xy$ plane $I$ is given by $I=l_{y}^{3}l_{z}/12$. When small asymmetry
is taken into account [see Eqs. (\ref{Y_0(theta) SA}) and (\ref{nu(theta) SA}%
)] the functions $f_{\mathrm{Y}}\left(  \theta\right)  $ and $\nu\left(
\theta\right)  $ can be expressed as (see Fig. \ref{Fig: f_y_and_f_nu})%
\begin{align}
f_{\mathrm{Y}}\left(  \theta\right)   &  =\operatorname{Re}\left(
\sqrt{\theta-\theta_{\mathrm{C}}-i\theta_{\mathrm{f}}}\right)  \;,
\label{f_Y}\\
f_{\nu}^{2}\left(  \theta\right)   &  =-\left(  \theta-\theta_{\mathrm{C}%
}\right)  +3f_{\mathrm{Y}}^{2}\left(  \theta\right)  \;, \label{f_nu}%
\end{align}
where $\theta_{\mathrm{C}}$ is the dimensionless buckling temperature, which
is related to the buckling temperature $T_{\mathrm{C}}$ by the relation
$\theta_{\mathrm{C}}=\left(  T_{\mathrm{C}}-T_{0}\right)  /T_{0}$, and where
the real dimensionless constant $\theta_{\mathrm{f}}$ represents the effect of asymmetry.

\section{Optical Cavity}

The finesse of the optical cavity is limited by loss mechanisms that give rise
to optical energy leaking out of the cavity. The main escape routes are
through the on fiber mirror (FBG or glass-vacuum interface), through
absorption by the metallic mirror, and through radiation; the corresponding
transmission probabilities are respectively denoted by $T_{\mathrm{B}}$,
$T_{\mathrm{A}}$ and $T_{\mathrm{R}}$. Let $y_{\mathrm{D}}$ be the
displacement of the mirror relative to a point, at which the energy stored in
the optical cavity in steady state obtains a local maximum. In the ideal case,
all optical properties of the cavity are periodic in $y_{\mathrm{D}}$ with a
period of $\lambda/2$, where $\lambda$ is the optical wavelength (though, as
can be seen in Fig. \ref{Fig:S_R}(a), deviation from periodic behavior is
experimentally observed).

It is assumed that $y_{\mathrm{D}}$ is related to $\xi$ by the following
relation $4\pi y_{\mathrm{D}}/\lambda=\beta_{\mathrm{Y}}\left(  \xi
+\xi_{\mathrm{f}}\right)  $, where both $\beta_{\mathrm{Y}}$ and
$\xi_{\mathrm{f}}$ are constants, which depend on the position of the fiber.
For a fixed $\xi$ the cavity reflection probability $R_{\mathrm{C}}$, i.e. the
ratio between the reflected (outgoing) and injected (incoming) optical powers
in the fiber, is given by \cite{Zaitsev_046605}%
\begin{equation}
R_{\mathrm{C}}=\frac{\left(  \frac{T_{\mathrm{B}}-T_{\mathrm{A}}%
-T_{\mathrm{R}}}{2}\right)  ^{2}+2\left[  1-\cos\left(  \beta_{\mathrm{Y}%
}\left(  \xi+\xi_{\mathrm{f}}\right)  \right)  \right]  }{\left(
\frac{T_{\mathrm{B}}+T_{\mathrm{A}}+T_{\mathrm{R}}}{2}\right)  ^{2}+2\left[
1-\cos\left(  \beta_{\mathrm{Y}}\left(  \xi+\xi_{\mathrm{f}}\right)  \right)
\right]  }\;. \label{eq:opt_power_Rc}%
\end{equation}
The heating power $Q$ due to optical absorption of the suspended
micromechanical mirror can be expressed as $Q=I\left(  \xi\right)
P_{\mathrm{L}}$, where $P_{\mathrm{L}}$ is the power of the monochromatic
laser light incident on the cavity and where the function $I\left(
\xi\right)  $ is given by \cite{Zaitsev_046605}%
\begin{equation}
I\left(  \xi\right)  =\frac{T_{\mathrm{B}}T_{\mathrm{A}}}{\left(
\frac{T_{\mathrm{B}}+T_{\mathrm{A}}+T_{\mathrm{R}}}{2}\right)  ^{2}+2\left[
1-\cos\left(  \beta_{\mathrm{Y}}\left(  \xi+\xi_{\mathrm{f}}\right)  \right)
\right]  }\;.
\end{equation}

\section{Thermal Balance Equation}

The temperature $T$ evolves according to the following thermal balance
equation%
\begin{equation}
\frac{\mathrm{d}T}{\mathrm{d}t}=\frac{Q}{C}-\frac{H\left(  T-T_{0}\right)
}{C}\;,
\end{equation}
where $C$ is heat capacity, $H$ is thermal transfer coefficient and $Q$ is the
heating power due to optical absorption. In terms of the dimensionless
temperature $\theta$ and dimensionless time $\tau$ the thermal balance
equation becomes%
\begin{equation}
\dot{\theta}=\beta_{\mathrm{P}}I\left(  \xi\right)  -\beta_{\mathrm{TR}}%
\theta\;, \label{eom theta}%
\end{equation}
where $\beta_{\mathrm{TR}}=\sqrt{\Upsilon/E}H/C$ is the dimensionless thermal
rate, and $\beta_{\mathrm{P}}=\sqrt{\Upsilon/E}P_{\mathrm{L}}/CT_{0}$ is the
dimensionless injected laser power.

\section{Small Amplitude Limit}

The coupling between the equation of motion for $\xi$ [Eq. (\ref{EOM xi})] and
the the one for $\theta$ [Eq. (\ref{eom theta})] origins by three terms, the
$\theta$ dependent frequency $\Omega$, the $\theta$ dependent force
$F_{\mathrm{th}}$ (i.e. the thermal force) [see Eq. (\ref{EOM xi})], and the
$\xi$ dependent optical heating power [the term $I\left(  \xi\right)  $ in Eq.
(\ref{eom theta})]. The case where the first two coupling terms are linearized
(i.e. approximated by a linear function of $\theta$) is identical to the case
that was studied in Ref. \cite{Zaitsev_1104_2235}, in which slow envelope
evolution equations for the system were derived, and the amplitudes and the
corresponding oscillation frequencies of different limit cycles were analyzed.
By employing the same analysis (and the same simplifying assumptions) for the
present case one finds that the coupling leads to renormalization of the
mechanical damping rate, which effectively becomes%
\begin{equation}
\gamma_{\mathrm{eff}}=\gamma+\frac{f_{\mathrm{Y}}^{\prime}I^{\prime}%
\xi_{\mathrm{C}}\beta_{\mathrm{P}}}{2\left(  1+\frac{\beta_{\mathrm{TR}}^{2}%
}{\Omega^{2}}\right)  }\;, \label{gamma_eff=}%
\end{equation}
where $I^{\prime}=\mathrm{d}I/\mathrm{d}\xi$ and where $f_{\mathrm{Y}}%
^{\prime}=\mathrm{d}f_{\mathrm{Y}}/\mathrm{d}\theta$. The corresponding
effective mechanical resonance frequency $\Omega_{\mathrm{eff}}$ is given by%
\begin{equation}
\Omega_{\mathrm{eff}}=\Omega-\frac{\left(  \gamma_{\mathrm{eff}}%
-\gamma\right)  \beta_{\mathrm{TR}}}{\Omega_{\mathrm{p}}}\;.
\label{Omega_eff=}%
\end{equation}

The white dotted line in Fig. \ref{Fig:S_R}(b) shows the calculated value of
$\Omega_{\mathrm{eff}}$ [see Eq. (\ref{Omega_eff=})] for the regions in
$V_{\mathrm{z}}$ for which $\gamma_{\mathrm{eff}}<0$ [see Eq.
(\ref{gamma_eff=})]. The system's parameters that were used for the
calculation are listed in the caption of Fig. \ref{Fig:S_R}. Note that Eqs.
(\ref{gamma_eff=}) and (\ref{Omega_eff=}), which were derived by assuming the
limit of small amplitudes, become inapplicable in most of the region, in which
self-excited oscillations are observed in Fig. \ref{Fig:S_R}(b), and
consequently, relatively large deviation between data and the prediction of
Eq. (\ref{Omega_eff=}) is expected.

The threshold of instability, i.e. Hopf bifurcation, occurs when
$\gamma_{\mathrm{eff}}$ vanishes. For the case where the dominant coupling
mechanism between the mechanical resonator and the optical field arises due to
radiation pressure, instability can occur only for the so-called case of blue
detuning, i.e. the case where $I^{\prime}<0$. However, contrary to the case of
radiation pressure \cite{Rodrigues_053601,Schliesser_243905}, which always
acts in the outwards direction, the thermal force due to buckling can act in
both directions. We refer to the case where $f_{\mathrm{Y}}^{\prime}>0$ as the
case of outwards buckling, in which due to buckling the optical cavity becomes
longer, whereas the opposite case of inwards buckling occurs when
$f_{\mathrm{Y}}^{\prime}<0$. As can be seen from Eq. (\ref{gamma_eff=}), the
threshold laser power $\beta_{\mathrm{P}}$ at which Hopf bifurcation occurs is
inversely proportional to $\left\vert f_{\mathrm{Y}}^{\prime}\right\vert $.
Close to the buckling temperature $\theta_{\mathrm{C}}$ and when asymmetry is
small (i.e. when $\left\vert \theta_{\mathrm{f}}\right\vert \ll1$) the factor
$\left\vert f_{\mathrm{Y}}^{\prime}\right\vert $ can become very large [see
Eq. (\ref{f_Y}) and Fig. \ref{Fig: f_y_and_f_nu}], and consequently, the
threshold laser power in that region can become very small.

\section{High Laser Power}

\begin{figure}
[ptb]
\begin{center}
\includegraphics[
height=2.4336in,
width=3.2396in
]%
{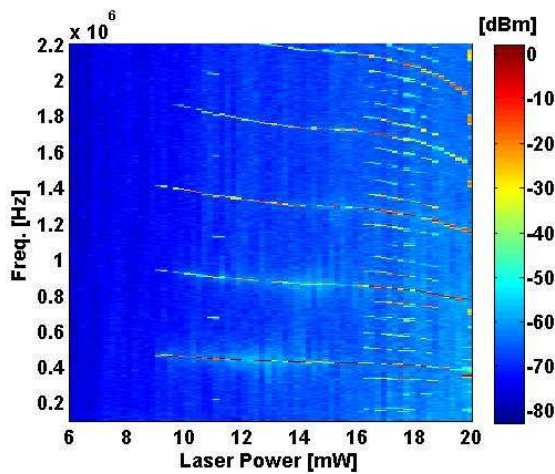}%
\caption{(Color Online) In addition to the harmonics in the measured spectrum
due to self-excited oscillations of mode 1 of beam B, subharmonics of order
$1/2$ and of order $1/5$ are observed.}%
\label{Fig. Y1}%
\end{center}
\end{figure}

\begin{figure}
[ptb]
\begin{center}
\includegraphics[
height=2.4336in,
width=3.2396in
]%
{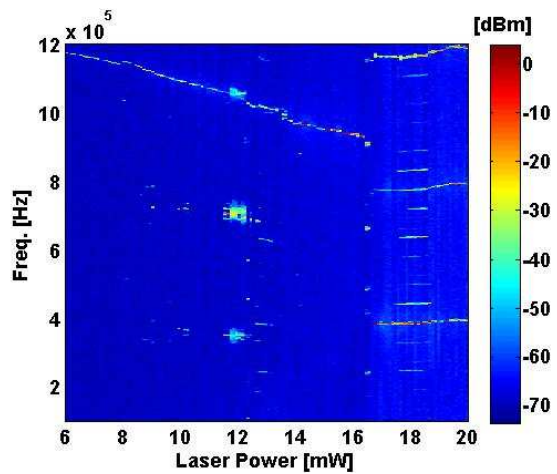}%
\caption{(Color Online) The subharmonics of order $1/3$ near input power of
$11.9\operatorname{mW}$ are attributed to excitation of mode 1 by the
self-oscillating mode 2, which has a frequency $f_{2}$ that is 3 times larger
than $f_{1}$ in that region. }%
\label{Fig. Y2}%
\end{center}
\end{figure}

Other interesting effects have been observed with relatively high laser power.
Two examples are presented below. The multilayer beam mirror that was used for
the experiments that are introduced in this section (see Figs. \ref{Fig. Y1}%
\ and \ref{Fig. Y2}) has the same layer structure as beam A [see Fig.
\ref{Fig:setup}(b)], however its length is $l=120%
\operatorname{\mu m}%
$ and its width is $b=20%
\operatorname{\mu m}%
$. We will refer to this beam hereafter as beam B. The same fabrication
process has been employed for both beams. The two lowest lying modes of bean
B, which are hereafter referred to as mode 1 and mode 2, have resonance
frequencies $f_{1}$ and $f_{2}$ of about $510%
\operatorname{kHz}%
$ and $1200%
\operatorname{kHz}%
$ respectively (note that these values vary with laser power and cavity length).%

Self-excited oscillations of mode 1 are seen in Fig. \ref{Fig. Y1}. In this
measurement the sample is kept grounded and the reflected optical power is
measured using a spectrum analyzer as a function of input optical power. In
the frequency window that has been employed for this measurement the first 5
harmonics of the self-oscillating mode 1 are seen. In addition, for some
values of input power the spectrum contains subharmonics. Close to input power
of $11%
\operatorname{mW}%
$ the subharmonics indicate that period doubling occurs, whereas in the range
$16.2-18.6%
\operatorname{mW}%
$ the period is multiplied by a factor of $5$. While the period doubling
effect has been predicted for a similar system by Blocher \textit{et al}.
\cite{Blocher_52}, the underlying mechanism responsible for the period
multiplication by a factor of $5$ is yet undetermined.

The data in Fig. \ref{Fig. Y2}, which was taken with a modified cavity length,
exhibits self-excited oscillations of mode 2 in the laser input power range
$6-16.5%
\operatorname{mW}%
$. Above that range self-excited oscillations of mode 1 are observed. Near
input power of $11.9%
\operatorname{mW}%
$ the ratio between $f_{2}$ and $f_{1}$ is $3$, i.e. $f_{2}=3f_{1}$. This
gives rise to strong intermode coupling, which leads to excitation of mode 1
by the self-oscillating mode 2, as can be seen from the subharmonics of order
$1/3$ that are measured in that range.

Note that for other cavity lengths the measured spectrum of reflected power
can exhibit wide band response when the input laser power is sufficiently
large. This possibly indicates that the dynamics becomes chaotic
\cite{Blocher_52}.%

\section{Summary}

In summary, an optomechanical cavity having a buckled mirror may exhibits some
unusual behaviors when the operating temperature is close to the buckling
temperature. In that range the instability threshold can be reached with
relatively low laser power. Moreover, contrary to the more common case, the
instability is obtained with red detuned laser light for the case of inwards
buckling. With relatively high laser power other interesting effects are
observed in the unstable region including period doubling and intermode
coupling. Further theoretical work is needed to account for such effects.

Both first authors (D. Y. and M. B. K.) have equally contributed to the paper.

\appendix

\section{Lagrangian}

Consider a beam made of a material having mass density $\rho$ and Young's
modulus $E$. In the absent of tension the length of the beam is $l_{0}$. The
beam is doubly clamped to a substrate at the points $\left(  x,y\right)
=\left(  \pm l/2,0\right)  $ and the motion of the beam's axis, which is
described by the height function $y\left(  x,t\right)  $, is assumed to be
exclusively in the $xy$ plane. The corresponding moment of inertia is denoted
by $I$. The Lagrangian functional $\mathcal{L}$ is given by
\cite{Lawrence_223}%
\begin{equation}
\mathcal{L}=\int\limits_{-l/2}^{l/2}\mathrm{d}x\;L-\frac{\beta_{\mathrm{A}}%
El}{8}\left(  \int\limits_{-l/2}^{l/2}\mathrm{d}x\left(  \frac{\partial
y}{\partial x}\right)  ^{2}\right)  ^{2}\;, \label{L beam}%
\end{equation}
where%
\begin{equation}
L=\frac{A_{\mathrm{cs}}\rho}{2}\left(  \frac{\partial y}{\partial t}\right)
^{2}-\frac{N}{2}\left(  \frac{\partial y}{\partial x}\right)  ^{2}-\frac
{EI}{2}\left(  \frac{\partial^{2}y}{\partial x^{2}}\right)  ^{2}+fy\;
\end{equation}
is the Lagrangian density, $A_{\mathrm{cs}}$ is the cross section area of the
beam and%
\begin{equation}
N=EA_{\mathrm{cs}}\frac{l-l_{0}}{l_{0}}%
\end{equation}
is the tension in the beam for the straight case where $y=0$. The beam's
equation of motion is found from the principle of least action to be given by
\cite{Nayfeh_1121}%
\begin{align}
\Upsilon\frac{\partial^{2}y}{\partial t^{2}}  &  =\left(  N+\frac
{A_{\mathrm{cs}}E}{2l}\int\limits_{-l/2}^{l/2}\mathrm{d}x\left(
\frac{\partial y}{\partial x}\right)  ^{2}\right)  \frac{\partial^{2}%
y}{\partial x^{2}}\nonumber\\
&  -EI\frac{\partial^{4}y}{\partial x^{4}}+f\;,\nonumber\\
&  \label{EOM V2}%
\end{align}
where $\Upsilon=\rho A_{\mathrm{cs}}$ is the mass density per unit length.

\section{First Buckling Configuration}

Consider the case where the deflection $y\left(  x,t\right)  $ has the shape
of the first buckling configuration \cite{Nayfeh_1121}, i.e. $y=\mathcal{Y}%
l\left(  1+\cos\frac{2\pi x}{l}\right)  $, where $\mathcal{Y}$ is a
dimensionless time dependent amplitude. In order to account for a possible
asymmetry, the force $f$ is allowed to be a nonzero constant
\cite{Lawrence_223,McCarthy_37}. The Lagrangian (\ref{L beam}) for the present
case becomes%
\begin{align}
\mathcal{L}_{0}  &  =\frac{3\beta_{\mathrm{A}}\rho l^{5}\left(  \frac
{\mathrm{d}\mathcal{Y}}{\mathrm{d}t}\right)  ^{2}}{4}-\pi^{2}Nl\mathcal{Y}%
^{2}\nonumber\\
&  -4\pi^{4}\beta_{\mathrm{I}}El^{3}\mathcal{Y}^{2}+fl^{2}\mathcal{Y}%
-\frac{\pi^{4}\beta_{\mathrm{A}}El^{3}\mathcal{Y}^{4}}{2}\;,\nonumber\\
&
\end{align}
where $\beta_{\mathrm{A}}=A_{\mathrm{cs}}/l^{2}$ and where $\beta_{\mathrm{I}%
}=I/l^{4}$. Alternatively, $\mathcal{L}_{0}$ can be expressed as%
\begin{equation}
\mathcal{L}_{0}=T_{0}-U_{0}\;,
\end{equation}
where the kinetic energy $T_{0}$ is given by%
\begin{equation}
T_{0}=\frac{m_{0}l^{2}\left(  \frac{\mathrm{d}\mathcal{Y}}{\mathrm{d}%
t}\right)  ^{2}}{2}\;,
\end{equation}
where $m_{0}=3\beta_{\mathrm{A}}\rho l^{3}/2$ and where the potential energy
$U_{0}$ is given by%
\begin{equation}
U_{0}=\frac{m_{0}\omega_{0}^{2}l^{2}}{2}u\left(  \mathcal{Y}\right)  \;,
\end{equation}
where%
\begin{equation}
u\left(  \mathcal{Y}\right)  =\eta_{1}\mathcal{Y}+\eta_{2}\mathcal{Y}^{2}%
+\eta_{4}\mathcal{Y}^{4}\;,
\end{equation}
$\eta_{1}=-2f/m_{0}\omega_{0}^{2}$, $\eta_{2}=1-\beta_{\mathrm{L}}$, $\eta
_{4}=\pi^{4}\beta_{\mathrm{A}}El/m_{0}\omega_{0}^{2}$, $\beta_{\mathrm{L}%
}=-N/4\pi^{2}\beta_{\mathrm{I}}El^{2}$ and where $\omega_{0}^{2}=\left(
2\pi\right)  ^{4}\beta_{\mathrm{I}}E/3\rho A_{\mathrm{cs}}$.

Local minima points $\mathcal{Y}_{0}$ of the potential $U_{0}$ are found by
solving $0=du/d\mathcal{Y}$. The dimensionless potential $u\left(
\mathcal{Y}\right)  $ can be expanded near one of its local minima points
$\mathcal{Y}_{0}$ to second order in $\mathcal{Y}-\mathcal{Y}_{0}\equiv\xi$ as
$u=u_{0}+\nu^{2}\xi^{2}+O\left(  \xi^{3}\right)  $, where both $u_{0}$ and
$\nu=\sqrt{\eta_{2}+6\eta_{4}\mathcal{Y}_{0}^{2}}$ are constants.

For the case where $\eta_{1}=0$ the coordinate $\mathcal{Y}_{0}$ is given by
$\mathcal{Y}_{0}=\pm\operatorname{Re}\sqrt{-\eta_{2}/2\eta_{4}}$. The solution
for the case of small $\left\vert \eta_{1}\right\vert $ can be approximated by%
\begin{equation}
\mathcal{Y}_{0}=\operatorname{Re}\sqrt{-\frac{\eta_{2}+i\left(  4\eta_{1}%
^{2}\eta_{4}\right)  ^{1/3}}{2\eta_{4}}}\;, \label{Y_0 SA}%
\end{equation}
provided that $\left\vert \eta_{2}\right\vert $ is small, i.e. provided that
the system is close to buckling conditions. The corresponding approximation
for $\nu$ is found from the relation $\nu=\sqrt{\eta_{2}+6\eta_{4}%
\mathcal{Y}_{0}^{2}}$.

\section{Mechanical equation of motion}

Let $\alpha$ and $\alpha_{\mathrm{s}}$ be the thermal expansion coefficients
of the metallic beam and of the substrate respectively. At some given
temperature $T_{\mathrm{TF}}$ the tension $N$ is assumed to vanish. Since the
substrate is much larger than the beam, one may assume that the substrate
thermally contracts (or expands) as if the suspended beam was not attached to
it, namely the distance between both clamping points at temperature
$T=T_{\mathrm{TF}}+\Delta_{\mathrm{T}}$ becomes $l_{\mathrm{s}}=l_{0}\left(
1+\alpha_{\mathrm{s}}\Delta_{\mathrm{T}}\right)  $, where $l_{0}$ is the
distance at the tension free temperature $T_{\mathrm{TF}}$. The tension $N$
needed to keep the beam clamped to the substrate is thus given by
$N=EA_{\mathrm{cs}}\left(  \alpha_{\mathrm{s}}-\alpha\right)  \Delta
_{\mathrm{T}}$ \cite{Pandey_203105}. Thus, the dimensionless parameter
$\beta_{\mathrm{L}}$ can be expressed as a function of the dimensionless
temperature%
\begin{equation}
\theta=\frac{T-T_{0}}{T_{0}}\;,
\end{equation}
where $T_{0}$ is the temperature of the supporting substrate (i.e. the base
temperature), as $\beta_{\mathrm{L}}=\beta_{\mathrm{A}}\beta_{\mathrm{T0}%
}\left(  \theta+\beta_{\mathrm{TF}}\right)  /4\pi^{2}\beta_{\mathrm{I}}$ where
$\beta_{\mathrm{T0}}=T_{0}\left(  \alpha-\alpha_{\mathrm{s}}\right)  $ and
where $\beta_{\mathrm{TF}}=\left(  T_{0}-T_{\mathrm{TF}}\right)  /T_{0}$.
Alternatively, $\beta_{\mathrm{L}}$ can be expressed as%
\begin{equation}
\beta_{\mathrm{L}}=1+\frac{\beta_{\mathrm{A}}\beta_{\mathrm{T0}}\left(
\theta-\theta_{\mathrm{C}}\right)  }{4\pi^{2}\beta_{\mathrm{I}}}\;,
\label{beta_L(theta)}%
\end{equation}
where $\theta_{\mathrm{C}}=4\pi^{2}\beta_{\mathrm{I}}/\beta_{\mathrm{A}}%
\beta_{\mathrm{T0}}-\beta_{\mathrm{TF}}$ is the dimensionless temperature at
which buckling occurs in the symmetric case (i.e. the case where $\eta_{1}=0$).

\begin{figure}
[ptb]
\begin{center}
\includegraphics[
height=2.4396in,
width=3.2396in
]%
{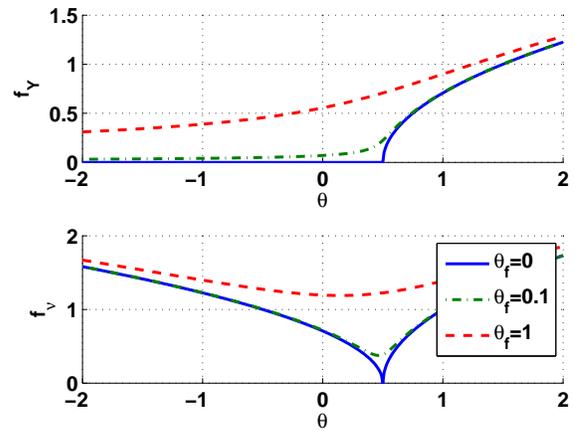}%
\caption{(Color Online) The functions $f_{\mathrm{Y}}\left(  \theta\right)  $
and $f_{\nu}\left(  \theta\right)  $ [see Eqs. (\ref{f_Y}) and (\ref{f_nu})]
plotted for 3 different values of the asymmetry parameter $\theta_{\mathrm{f}%
}$ and for the case where the dimensionless buckling temperature is
$\theta_{\mathrm{C}}=0.5$.}%
\label{Fig: f_y_and_f_nu}%
\end{center}
\end{figure}

With the help of Eq. (\ref{beta_L(theta)}) the parameters $\mathcal{Y}_{0}$
and $\nu$ can be expressed as a function of the dimensionless temperature
$\theta$. For the case of small asymmetry one has [see Eq. (\ref{Y_0 SA})]%
\begin{equation}
\mathcal{Y}_{0}=\xi_{\mathrm{C}}f_{\mathrm{Y}}\left(  \theta\right)  \;,
\label{Y_0(theta) SA}%
\end{equation}
where the function $f_{\mathrm{Y}}\left(  \theta\right)  $ is given by Eq.
(\ref{f_Y}), $\theta_{\mathrm{f}}=4\pi^{2}\beta_{\mathrm{I}}\left(  4\eta
_{1}^{2}\eta_{4}\right)  ^{1/3}/\beta_{\mathrm{A}}\beta_{\mathrm{T0}}$ and the
coefficient $\xi_{\mathrm{C}}$ is given by%
\begin{equation}
\xi_{\mathrm{C}}=\sqrt{\frac{\beta_{\mathrm{A}}\beta_{\mathrm{T0}}}{8\pi
^{2}\beta_{\mathrm{I}}\eta_{4}}}=\sqrt{\frac{3T_{0}\left(  \alpha
-\alpha_{\mathrm{s}}\right)  A_{\mathrm{cs}}\rho l^{4}\omega_{0}^{2}}%
{16\pi^{6}EI}}\;. \label{Y_0C}%
\end{equation}
The corresponding approximation for $\nu$ is given by%
\begin{equation}
\nu=\nu_{\mathrm{C}}f_{\nu}\left(  \theta\right)  \;, \label{nu(theta) SA}%
\end{equation}
where%
\begin{equation}
\nu_{\mathrm{C}}=\sqrt{\frac{\beta_{\mathrm{A}}\beta_{\mathrm{T0}}}{4\pi
^{2}\beta_{\mathrm{I}}}}=\sqrt{\frac{A_{\mathrm{cs}}l^{2}T_{0}\left(
\alpha-\alpha_{\mathrm{s}}\right)  }{4\pi^{2}I}}\;,
\end{equation}
and where the function $f_{\nu}\left(  \theta\right)  $ is given by Eq.
(\ref{f_nu}). The functions $f_{\mathrm{Y}}\left(  \theta\right)  $ and
$f_{\nu}\left(  \theta\right)  $ are plotted in Fig. \ref{Fig: f_y_and_f_nu}.%

The temperature dependence of $\mathcal{Y}_{0}$ can be taken into account by
adding a thermal force term $F_{\mathrm{th}}$ \cite{Fang_116, Fang&Wickert_96,
Leong_et_al_08} to the equation of motion of the mechanical amplitude
$\xi=\mathcal{Y}-\mathcal{Y}_{0}$. Similarly, the temperature dependence of
$\nu$ can be taken into account by including a thermal frequency shift term.
When in addition damping and external driving are taken into account the
equation of motion (\ref{EOM xi}) for $\xi$ is obtained.


\section*{Acknowledgment}

We would like to thank Yuli Starosvetsky for many fruitful discussions and
important comments. This work was supported by the German Israel Foundation
under Grant No. 1-2038.1114.07, the Israel Science Foundation under Grant No.
1380021, the Deborah Foundation, the Mitchel Foundation, Ministry of Science,
Russell Berrie Nanotechnology Institute, the European STREP QNEMS Project,
MAGNET Metro 450 consortium and MAFAT.

\newpage
\bibliographystyle{ieee}
\bibliography{acompat,Eyal_Bib}

\newif\ifabfull\abfulltrue
\begin{thebibliography}{10}

\bibitem{Braginsky&Manukin_67}
V.~B. Braginsky and A.~B. Manukin,
\newblock ``Ponderomotive effects of electromagnetic radiation (in
  {R}ussian),''
\newblock {\em ZhETF}, vol. 52, pp. 986--989, 1967.

\bibitem{Gigan_67}
S.~Gigan, H.~R. B{\"o}hm, M.~Paternostro, F.~Blaser, J.~B. Hertzberg, K.~C.
  Schwab, D.~Bauerle, M.~Aspelmeyer, and A.Zeilinger,
\newblock ``Self cooling of a micromirror by radiation pressure,''
\newblock {\em Nature}, vol. 444, pp. 67--70, 2006.

\bibitem{Metzger_1002}
C.~H. Metzger and K.Karrai,
\newblock ``Cavity cooling of a microlever,''
\newblock {\em Nature}, vol. 432, pp. 1002--1005, 2004.

\bibitem{Kippenberg_1172}
T.~J. Kippenberg and K.~J. Vahala,
\newblock ``Cavity optomechanics: Back-action at the mesoscale,''
\newblock {\em Science}, vol. 321, no. 5893, pp. 1172--1176, Aug 2008.

\bibitem{Favero_104101}
C.~Metzger I.~Favero, S.~Camerer, D.~Konig, H.~Lorenz, J.~P. Kotthaus, and
  K.~Karrai,
\newblock ``Optical cooling of a micromirror of wavelength size,''
\newblock {\em Appl. Phys. Lett.}, vol. 90, pp. 104101, 2007.

\bibitem{Hane_179}
K.~Hane and K.~Suzuki,
\newblock ``Self-excited vibration of a self-supporting thin film caused by
  laser irradiation,''
\newblock {\em Sensors and Actuators A: Physical}, vol. 51, pp. 179--182, 1996.

\bibitem{Kimble_et_al_01}
H.~J. Kimble, Y.~Levin, A.~B. Matsko, K.~S. Thorne, and S.~P. Vyatchanin,
\newblock ``Conversion of conventional gravitational-wave interferometers into
  quantum nondemolition interferometers by modifying their input and/or output
  optics,''
\newblock {\em Phys. Rev. D}, vol. 65, pp. 022002, Dec 2001.

\bibitem{Carmon_et_al_05}
T.~Carmon, H.~Rokhsari, L.~Yang, T.~J. Kippenberg, and K.~J. Vahala,
\newblock ``Temporal behavior of radiation-pressure-induced vibrations of an
  optical microcavity phonon mode,''
\newblock {\em Phys. Rev. Lett.}, vol. 94, pp. 223902, Jun 2005.

\bibitem{Arcizet_et_al_06}
O.~Arcizet, P.-F. Cohadon, T.~Briant, M.~Pinard, and A.~Heidmann,
\newblock ``Radiation-pressure cooling and optomechanical instability of a
  micromirror,''
\newblock {\em Nature}, vol. 444, pp. 71--74, Nov 2006.

\bibitem{Jayich_et_al_08}
A.~M. Jayich, J.~C. Sankey, B.~M. Zwickl, C.~Yang, J.~D. Thompson, S.~M.
  Girvin, A.~A. Clerk, F.~Marquardt, and J.~G.~E. Harris,
\newblock ``Dispersive optomechanics: a membrane inside a cavity,''
\newblock {\em New J. Phys.}, vol. 10, pp. 095008, Sep 2008.

\bibitem{Schliesser_et_al_08}
A.~Schliesser, R.~Riviere, G.~Anetsberger, O.~Arcizet, and T.~J. Kippenberg,
\newblock ``Resolved-sideband cooling of a micromechanical oscillator,''
\newblock {\em Nat. Phys.}, vol. 4, pp. 415--419, 2008.

\bibitem{Genes_et_al_08}
C.~Genes, D.~Vitali, P.~Tombesi, S.~Gigan, and M.~Aspelmeyer,
\newblock ``Ground-state cooling of a micromechanical oscillator: {C}omparing
  cold damping and cavity-assisted cooling schemes,''
\newblock {\em Phys. Rev. A}, vol. 77, pp. 033804, Mar 2008.

\bibitem{Teufel_et_al_10}
J.~D. Teufel, D.~Li, M.~S. Allman, K.~Cicak, A.~J. Sirois, J.~D. Whittaker, and
  R.~W. Simmonds,
\newblock ``Circuit cavity electromechanics in the strong coupling regime,''
\newblock {\em arXiv}, Nov 2010.

\bibitem{Poot_273}
Menno Poot and Herre~S.J. van~der Zant,
\newblock ``Mechanical systems in the quantum regime,''
\newblock {\em Phys. Rep.}, vol. 511, pp. 273--335, 2012.

\bibitem{Wu_et_al_06}
M.~C. Wu, O.~Solgaard, and J.~E. Ford,
\newblock ``Optical {MEMS} for lightwave communication,''
\newblock {\em J. Lightwave Technol.}, vol. 24, no. 12, pp. 4433--4454, Dec
  2006.

\bibitem{Stokes_et_al_90}
N.~A.~D. Stokes, R.~M.~A. Fatah, and S.~Venkatesh,
\newblock ``Self-excitation in fibre-optic microresonator sensors,''
\newblock {\em Sens. Actuators, A}, vol. 21, pp. 369--372, Feb 1990.

\bibitem{Lyshevski&Lyshevski_03}
S.E. Lyshevski and M.A. Lyshevski,
\newblock ``Nano- and microoptoelectromechanical systems and nanoscale active
  optics,''
\newblock in {\em Third IEEE Conference on Nanotechnology, 2003.}, Aug 2003,
  vol.~2, pp. 840--843.

\bibitem{Hossein-Zadeh&Vahala_10}
M.~Hossein-Zadeh and K.~J. Vahala,
\newblock ``An optomechanical oscillator on a silicon chip,''
\newblock {\em IEEE J. Sel. Top. Quantum Electron.}, vol. 16, no. 1, pp.
  276--287, Jan 2010.

\bibitem{Jourdan_et_al_08}
G.~Jourdan, F.~Comin, and J.~Chevrier,
\newblock ``Mechanical mode dependence of bolometric backaction in an atomic
  force microscopy microlever,''
\newblock {\em Phys. Rev. Lett.}, vol. 101, pp. 133904, Sep 2008.

\bibitem{Marino&Marin_10}
F.~Marino and F.~Marin,
\newblock ``Chaotically spiking attractors in suspended mirror optical
  cavities,''
\newblock {\em arXiv}, Jun 2010.

\bibitem{Liberato_et_al_10}
S.~D. Liberato, N.~Lambert, and F.~Nori,
\newblock ``Quantum limit of photothermal cooling,''
\newblock {\em arXiv}, Nov 2010.

\bibitem{Paternostro_et_al_06}
M.~Paternostro, S.~Gigan, M.~S. Kim, F.~Blaser, H.~R. B{\"o}hm, and
  M.~Aspelmeyer,
\newblock ``Reconstructing the dynamics of a movable mirror in a detuned
  optical cavity,''
\newblock {\em New J. Phys.}, vol. 8, pp. 107, Jun 2006.

\bibitem{Marquardt_103901}
F.~Marquardt, J.~G.~E. Harris, and S.~M. Girvin,
\newblock ,''
\newblock {\em Phys. Rev. Lett.}, vol. 96, pp. 103901, 2006.

\bibitem{Metzger_133903}
C.~Metzger, M.~Ludwig, C.~Neuenhahn, A.~Ortlieb, I.~Favero, K.~Karrai, and
  F.~Marquardt,
\newblock ``Self-induced oscillations in an optomechanical system driven by
  bolometric backaction,''
\newblock {\em Phys. Rev. Lett.}, vol. 101, pp. 133903, Sep 2008.

\bibitem{Restrepo_860}
J.~Restrepo, J.~Gabelli, C.~Ciuti, and I.~Favero,
\newblock ``Classical and quantum theory of photothermal cavity cooling of a
  mechanical oscillator,''
\newblock {\em Comptes Rendus Physique}, vol. 12, pp. 860--870, Nov 2011.

\bibitem{Aubin_et_al_04}
K.~Aubin, M.~Zalalutdinov, T.~Alan, R.B. Reichenbach, R.~Rand, A.~Zehnder,
  J.~Parpia, and H.~Craighead,
\newblock ``Limit cycle oscillations in {CW} laser-driven {NEMS},''
\newblock {\em J. Microelectromech. Syst.}, vol. 13, pp. 1018 -- 1026, Dec
  2004.

\bibitem{Kim_1454225}
Kiwoong Kim and Soonchil Lee,
\newblock ``Self-oscillation mode induced in an atomic force microscope
  cantilever,''
\newblock {\em J. Appl. Phys. 91, 4715 (2002); doi:10.1063/1.1454225 (5 pages)
  Self-oscillation mode induced in an atomic force microscope cantilever
  Kiwoong Kim and Soonchil Lee}, vol. J. Appl. Phys., pp. 1454225, 2002.

\bibitem{Aubin_1018}
K.~Aubin, M.~Zalalutdinov, T.~Alan, R.B. Reichenbach, R.~Rand, A.~Zehnder,
  J.~Parpia, and H.~Craighead,
\newblock ``Limit cycle oscillations in {CW} laser-driven {NEMS},''
\newblock {\em J. MEMS}, vol. 13, pp. 1018--1026, 2004.

\bibitem{Carmon_223902}
T.~Carmon, H.~Rokhsari, L.~Yang, T.~J. Kippenberg, and K.~J. Vahala,
\newblock ,''
\newblock {\em Phys. Rev. Lett.}, vol. 94, pp. 223902, 2005.

\bibitem{Corbitt_021802}
T.~Corbitt, D.~Ottaway, E.~Innerhofer, J.~Pelc, and N.~Mavalvala,
\newblock ,''
\newblock {\em Phys. Rev. A}, vol. 74, pp. 21802, 2006.

\bibitem{Carmon_123901}
T.~Carmon and K.~J. Vahala,
\newblock ,''
\newblock {\em Phys. Rev. Lett.}, vol. 98, pp. 123901, 2007.

\bibitem{Zaitsev_1104_2235}
Stav Zaitsev, Oded Gottlieb, and Eyal Buks,
\newblock ``Nonlinear dynamics of a microelectromechanical mirror in an optical
  resonance cavity,''
\newblock {\em Nonlinear Dynamics (in press), arXiv:1104.2235}, 2011.

\bibitem{Zaitsev_046605}
Stav Zaitsev, Ashok~K. Pandey, Oleg Shtempluck, and Eyal Buks,
\newblock ``Forced and self-excited oscillations of optomechanical cavity,''
\newblock {\em Phys. Rev. E 84}, vol. 84, pp. 046605, 2011.

\bibitem{Riethmuller_758}
W.~Riethmuller and W.~Benecke,
\newblock ``Thermally excited silicon microactuators,''
\newblock {\em IEEE Trans. Electron. Dev.}, vol. 35, pp. 758--762, 1988.

\bibitem{Michael_870}
A.~Michael and C.Y. Kwok,
\newblock ``Buckling shape of elastically constrained multi-layered
  micro-bridges,''
\newblock {\em Sensors and Actuators A}, vol. 135, pp. 870--880, 2007.

\bibitem{Ross_537}
David~S. Ross, Antonio Cabal, David Trauernicht, and John Lebens,
\newblock ``Temperature-dependent vibrations of bilayer microbeams,''
\newblock {\em Sensors and Actuators A}, vol. 119, pp. 53--543, 2005.

\bibitem{Rodrigues_053601}
D.~A. Rodrigues and A.~D. Armour,
\newblock ``Amplitude noise suppression in cavity-driven oscillations of a
  mechanical resonator,''
\newblock {\em Phys. Rev. Lett.}, vol. 104, pp. 053601, Feb 2010.

\bibitem{Okamoto_062202}
Hajime Okamoto, Takehito Kamada, Koji Onomitsu, Imran Mahboob, and Hiroshi
  Yamaguchi,
\newblock ,''
\newblock {\em Applied Physics Express}, vol. 2, pp. 062202, 2009.

\bibitem{Coffin_935}
Douglas~W. Coffin and Frederick Bloom,
\newblock ``Elastica solution for the hygrothermal buckling of a beam,''
\newblock {\em International Journal of Non-Linear Mechanics}, vol. 34, pp.
  935--947, 1999.

\bibitem{Fang_116}
W.~Fang and J.A. Wickert,
\newblock ``Post buckling of micromachined beams,''
\newblock {\em J. Micromech. Microeng.}, vol. 4, pp. 116--122, 1994.

\bibitem{Gere_Mechanics_of_Mat}
J.~M. Gere and S.~P. Timoshenko,
\newblock {\em Mechanics of Materials},
\newblock PWS, Boston, MA, 1997.

\bibitem{Talghader_R109}
Joseph~J Talghader,
\newblock ``Thermal and mechanical phenomena in micromechanical optics,''
\newblock {\em J. Phys. D: Appl. Phys.}, vol. 37, pp. R109--R122, 2004.

\bibitem{Ettouhami_167}
A.~Ettouhami, A.~Essaid, N.~Ouakrim, L.~Michel, and M.~Limou.D,
\newblock ``Thermal buckling of silicon capacitive pressure sensor,''
\newblock {\em Sensors and Actuators A}, vol. 57, pp. 167--171, 1996.

\bibitem{McCarthy_37}
Matthew McCarthy, Nicholas Tiliakos, Vijay Modi, and Luc~G. Frechette,
\newblock ``Thermal buckling of eccentric microfabricated nickel beams as
  temperature regulated nonlinear actuators for flow control,''
\newblock {\em Sensors and Actuators A: Physical}, vol. 28, pp. 37--46, 2007.

\bibitem{Bagheri_726}
Mahmood Bagheri, Menno Poot, Mo~Li, Wolfram P.~H. Pernice, and Hong~X. Tang,
\newblock ``Dynamic manipulation of nanomechanical resonators in the
  high-amplitude regime and non-volatile mechanical memory operation,''
\newblock {\em Nature Nanotechnology}, vol. 6, pp. 726--732, 2011.

\bibitem{Roodenburg_183501}
D.~Roodenburg, J.~W. Spronck, H.~S.~J. van~der Zant, and W.~J. Venstra,
\newblock ``Buckling beam micromechanical memory with on-chip readout,''
\newblock {\em Appl. Phys. Lett.}, vol. 94, pp. 183501, 2009.

\bibitem{Nayfeh_1121}
A.~H. {Nayfeh}, W.~{Kreider}, and T.~J. {Anderson},
\newblock ``{Investigation of natural frequencies and mode shapes of buckled
  beams},''
\newblock {\em AIAA Journal}, vol. 33, pp. 1121--1126, June 1995.

\bibitem{Lawrence_223}
W.~E. Lawrence, M.~N. Wybourne, and S.~M. Carr,
\newblock ``Compressional mode softening and euler buckling patterns in
  mesoscopic beams,''
\newblock {\em N. J. Phys.}, vol. 8, pp. 223, 2006.

\bibitem{Blocher_52}
David Blocher, Alan~T. Zehnder, Richard~H. Rand, and Shreyasi Mukerji,
\newblock ``Anchor deformations drive limit cycle oscillations in
  interferometrically transduced mems beams,''
\newblock {\em Finite Elem. Anal. Des.}, vol. 49, pp. 52--57, Feb. 2012.

\bibitem{Snyder&Love_book_83}
A.~W. Snyder and J.~D. Love,
\newblock {\em Optical waveguide theory},
\newblock Springer, 1983.

\bibitem{Snyder_Optical_Waveguide_Th}
A.~W. Snyder and J.~D. Love,
\newblock {\em Optical Waveguide Theory},
\newblock Chapman and Hall, 1983.

\bibitem{Poladian_2963}
"Resonance Mode~Expansions L.~Poladian and 2963—2975~(1996). Exact
  Solutions~for Nonuniform~Gratings, " Phys. Rev. E~54,
\newblock ``L. poladian,''
\newblock {\em Phys. Rev. E}, vol. 54, pp. 2963--2975, 1996.

\bibitem{Mao_5887}
Youxin Mao, Shoude Chang, Sherif Sherif, and Costel Flueraru,
\newblock ``Graded-index fiber lens proposed for ultrasmall probes used in
  biomedical imaging,''
\newblock {\em Appl. Opt.}, vol. 46, pp. 5887--5894, 2007.

\bibitem{Schliesser_243905}
A.~Schliesser, P.~Del'Haye, N.~Nooshi, K.~J. Vahala, and T.~J. Kippenberg,
\newblock ``Radiation pressure cooling of a micromechanical oscillator using
  dynamical backaction,''
\newblock {\em Phys. Rev. Lett.}, vol. 97, pp. 243905, 2006.

\bibitem{Pandey_203105}
Ashok~Kumar Pandey, Oded Gottlieb, Oleg Shtempluck, and Eyal Buks,
\newblock ``Performance of an {AuPd} micromechanical resonator as a temperature
  sensor,''
\newblock {\em Appl. Phys. Lett.}, vol. 96, pp. 203105, 2010.

\bibitem{Fang&Wickert_96}
W.~Fang and J.~A. Wickert,
\newblock ``Determining mean and gradient residual stresses in thin films using
  micromachined cantilevers,''
\newblock {\em J. Micromech. Microeng.}, vol. 6, pp. 301--309, 1996.

\bibitem{Leong_et_al_08}
T.~G. Leong, B.~R. Benson, E.~K. Call, and D.~H. Gracias,
\newblock ``Thin film stress driven self-folding of microstructured
  containers,''
\newblock {\em Small}, vol. 4, no. 10, pp. 1605--1609, Oct 2008.

\end{thebibliography}

\end{document}